\documentclass[prl,%preprint,tightenlines,%
superscriptaddress,showpacs,floatfix,nofootinbib,twocolumn ]{revtex4}
\usepackage{graphics}
\usepackage{epsfig}
\begin{document}
\preprint{MKPH-T-04-22}
\title{Quantum electrodynamics for vector mesons}
\author{Dalibor Djukanovic}
\affiliation{Institut f\"ur Kernphysik, Johannes Gutenberg-Universit\"at, 55099
Mainz, Germany}
\author{Matthias R.~Schindler}
\affiliation{Institut f\"ur Kernphysik, Johannes Gutenberg-Universit\"at, 55099
Mainz, Germany}
\author{Jambul Gegelia}
\affiliation{Institut f\"ur Kernphysik, Johannes Gutenberg-Universit\"at, 55099
Mainz, Germany} \affiliation{High Energy Physics Institute, Tbilisi State University,
Tbilisi, Georgia}
\author{Stefan Scherer}
\affiliation{Institut f\"ur Kernphysik, Johannes Gutenberg-Universit\"at, 55099
Mainz, Germany}
\begin{abstract}
   Quantum electrodynamics for $\rho$ mesons is considered.
It is shown that, at tree level, the value of the gyromagnetic
ratio of the $\rho^+$ is fixed to 2 in a self-consistent effective
quantum field theory. Further, the mixing parameter of the photon
and the neutral vector meson is equal to the ratio of
electromagnetic and strong couplings, leading to the mass
difference $M_{\rho^0}-M_{\rho^\pm}\sim 1 \ {\rm MeV}$ at tree
order.
\end{abstract}
\date{December 10, 2004}
\maketitle

    The question of the intrinsic magnetic moment of (elementary) particles
of arbitrary spin $s$ has been discussed controversially in the literature and is
still of great interest.
    On the one hand, low-energy theorems and the optical theorem require that
the gyromagnetic ratio $g\simeq 2$ for a particle with arbitrary spin $s$
different from zero (at least for particles which do not participate in strong
interactions) \cite{weinbergg2}.
   On the other hand, general arguments have been given that the minimal coupling leads
to $1/s$ for this quantity \cite{Hagen:1970wn}.
   Finally, the investigations of Ref.~\cite{Lee:1962vm} regarding the theory of charged
vector mesons interacting with the electromagnetic field suggested that the
gyromagnetic ratio depends on a free parameter, thus allowing it to take any
value.
   Below, we will address this question from the point of view of effective
field theory (EFT).

   In Ref.\ \cite{Djukanovic:2004mm} we have shown how the universal
coupling of the $\rho$ meson and the Kawarabayashi-Suzuki-Riadzuddin-Fayyazuddin
(KSRF) relation \cite{Kawarabayashi:1966kd,Riazuddin:sw} follow from the
requirement that chiral perturbation theory of pions, nucleons, and $\rho$ mesons
is a consistent EFT.
   Although EFTs are non-renormalizable
in the traditional sense, the general principles of EFT \cite{Weinberg:1978kz}
require that all ultraviolet divergences can be absorbed into the redefinition of
fields and parameters of the most general Lagrangian \cite{Weinberg:mt}.
   Imposing the renormalizability in this sense one finds that not all
parameters of the most general Lagrangian are free but satisfy consistency
conditions \cite{ren}.
   In this Letter we will use similar arguments for the
effective Lagrangian including, in addition, the interaction with
photons to show that the gyromagnetic ratio is fixed to $g=2$ at
tree level.
    Furthermore, the mixing parameter of the photon and the neutral
vector meson is also fixed and leads to $M_{\rho^0}-M_{\rho^\pm}\sim 1 \ {\rm
MeV}$ at tree order.

   We start with the chirally invariant effective Lagrangian
including vector mesons in the form given by Weinberg \cite{Weinberg:de},
containing {\em all} interaction terms which respect Lorentz invariance, the
discrete symmetries, and chiral symmetry.
   The electromagnetic interaction is introduced by adding all terms with photon
fields which are allowed by U(1) gauge invariance,
\begin{eqnarray}
{\cal L}&=&-\frac{1}{4} \  B_{\mu\nu}B^{\mu\nu}-\frac{1}{4} \
 G^a_{\mu\nu}G^{a \mu\nu}+\frac{M_0^2}{2} \ V_{\mu}^a V^{a
\mu}\nonumber \\
&&+\frac{c_0}{2} \ B^{\mu\nu} V^3_{\mu\nu}+\frac{\kappa_0}{2} \ \epsilon^{3ab}
B^{\mu\nu} V^a_\mu V^b_\nu
\nonumber \\
&&+i\bar\Psi\gamma^\mu\left(\partial_\mu+ ie_0\frac{1+\tau^3}{2} B_\mu\right)
\Psi\nonumber\\ &&-m_0 \bar\Psi \Psi +g_0 \bar\Psi\gamma^\mu \frac{{\tau}^a}{2}
\Psi V^a_\mu + {\cal L}_1. \label{lagr}
\end{eqnarray}
   Here, $B_\mu$ is a U(1) gauge vector field, $V^a_{\mu}$ ($a=1,2,3$)  denote the
Cartesian components of an isospin triplet of vector fields, and $\Psi$ is an
isospin doublet of nucleon fields with mass $m_0$ \cite{f1}.
   Furthermore,
\begin{eqnarray*}
B_{\mu\nu}&=&\partial_\mu B_{\nu}-\partial_\nu B_{\mu},\nonumber \\
V^a_{\mu\nu}&=&\partial_\mu V^a_{\nu}-\partial_\nu V^a_{\mu},\nonumber\\
G^a_{\mu\nu}&=& V^a_{\mu\nu}+g_0 \ \epsilon^{abc} V_{\mu}^b V_{\nu}^c + e_0 \
\epsilon^{3ab} ( B_{\mu} V^b_{\nu} - B_{\nu} V^b_{\mu}). \label{def}
\end{eqnarray*}
   All fields and coupling constants in Eq.\ (\ref{lagr}) are bare quantities.
   From the point of view of EFT it is not consistent to
consider a minimal coupling only (see, e.g., Ref.~\cite{Koch:2001ii}).
   Using symmetry arguments only, $c_0$ and $\kappa_0$ are free parameters
of the most general effective Lagrangian \cite{kappa}.
   They contribute to the mixing of the photon and the neutral vector meson, and
to the magnetic (and quadrupole) moment of the charged vector mesons
\cite{Lee:1962vm}, respectively.
   Finally, ${\cal L}_1$ contains an infinite number of interaction terms which
are allowed by symmetries \cite{Weinberg:1978kz,Weinberg:mt}.

   Below we perform a one-loop order analysis of the $\rho$-meson
self-energy and the  $\rho\bar\psi\psi$ vertex functions.
   To that end, we first introduce renormalized fields $A_\mu$,
$\rho_\mu^0$, $\rho_\mu^\pm$, and $\psi$ as
\begin{eqnarray}
B_\mu&=&\sqrt{Z_A} A_\mu+\delta\lambda \ \rho^0_\mu, \nonumber\\
V^3_\mu&=&\sqrt{Z_0} \ \rho^0_\mu, \nonumber  \\
\frac{V^1_\mu \mp i V^2_\mu}{\sqrt{2}}&=& \sqrt{Z_\pm}\ \rho^\pm_\mu,
\nonumber\\
\Psi&=&\sqrt{Z_\Psi}\psi,
% \ \ \ \bar\Psi =\bar\psi \barZ_\Psi^{1/2},
\label{chvr}
\end{eqnarray}
with wave-function renormalization constants
\begin{eqnarray}
Z_A&=&1+\delta Z_A,\nonumber\\
Z_0&=&1+\delta Z_0,\nonumber\\
Z_\pm&=&1+\delta Z_\pm,\nonumber\\
Z_\Psi&=&1+\delta Z_\Psi. \label{Z}
\end{eqnarray}
   The function $\delta\lambda$ allows for a linear superposition of
   the neutral vector fields.
Note that the wave-function renormalization constants for
   the neutral $\rho$ meson, $Z_0$, and for the charged $\rho$
mesons, $Z_\pm$, differ from each other.
   Finally, $\sqrt{Z_\Psi}$ is a real diagonal $2\times 2$ matrix.
   For the renormalization of the masses and coupling constants we write
\begin{eqnarray}
c_0&=&c+\delta c,\nonumber\\
g_0&=&g_s+\delta g,\nonumber\\
e_0&=&e+\delta e,\nonumber\\
\kappa_0&=&\kappa +\delta\kappa,\nonumber\\
M_0^2&=&M^2+\delta M^2,\nonumber\\
m_0&=&m+\delta m, \label{spell}
\end{eqnarray}
   where the functions $\delta c$ {\em etc.} depend on all renormalized
coupling constants (and the renormalization prescription).
   Using Eqs.~(\ref{chvr}) - (\ref{spell}) we rewrite the Lagrangian of
Eq.\ (\ref{lagr}) as the sum of the basic Lagrangian, the counterterm Lagrangian,
and a remainder \cite{Collins:xc}:
\begin{displaymath}
{\cal L}={\cal L}_{\rm basic}+{\cal L}_{\rm ct}+\tilde{\cal L}_1.
\end{displaymath}
   The basic Lagrangian is given by
\begin{eqnarray}
{\cal L}_{\rm basic}&=&-\frac{1}{4} \ F_{\mu\nu}F^{\mu\nu}\nonumber\\
&& -\frac{1}{4} \ \rho^0_{\mu\nu} \rho^{0 \mu\nu}+\frac{M^2}{2} \ \rho_\mu^0
\rho^{0 \mu} +\frac{c}{2} \ F^{\mu\nu}
\tilde\rho^0_{\mu\nu}\nonumber\\
&& -\frac{1}{2} \ \rho^+_{\mu\nu}\rho^{- \mu\nu} +M^2 \ \rho_\mu^+ \rho^{- \mu}
-i \kappa \ F^{\mu\nu} \rho^+_\mu \rho^-_\nu\nonumber\\
&&
  + i\bar\psi \gamma^\mu \left(
\partial_\mu+ie \frac{1+\tau^0}{2} A_\mu\right)
\psi -m \bar\psi\psi \nonumber \\
&& +g_s \bar\psi \gamma^\mu \ \frac{{\tau}^\alpha}{2} \
 \psi \ \rho^\alpha_\mu,
 \label{lagrmipctl}
\end{eqnarray}
with the field-strength tensors
\begin{eqnarray*}
F_{\mu\nu}&=&\partial_\mu A_\nu-\partial_\nu A_\mu,\\
\rho^0_{\mu\nu}&=&\tilde{\rho}^0_{\mu\nu}-i g_s(\rho^+_\mu
\rho^-_\nu-\rho^-_\mu\rho^+_\nu),\\
\rho^\pm_{\mu\nu}&=&\tilde{\rho}^\pm_{\mu\nu}\mp i g_s(\rho^0_\mu
\rho^\pm_\nu-\rho^0_\nu \rho^\pm_\mu)\pm ie(A_\mu\rho^\pm_\nu-A_\nu\rho^\pm_\mu),
\end{eqnarray*}
and the auxiliary quantity $\tilde{\rho}^\alpha_{\mu\nu}=\partial_\mu
\rho^\alpha_\nu-\partial_\nu \rho^\alpha_\mu$, $\alpha=\pm,0$.
   In Eq.\ (\ref{lagrmipctl}), we introduced $\tau^{\pm}=\left(\tau^1\pm i\tau^2\right)/\sqrt{2}$
and $\tau^0=\tau^3$.
   The counterterm Lagrangian (at one-loop order) is given by
\begin{eqnarray}
{\cal L}_{\rm ct}&=&-\frac{\delta Z_0-2 c \delta\lambda}{4} \
\tilde\rho^0_{\mu\nu} \tilde\rho^{0 \mu\nu} +\frac{\delta M^2+M^2 \delta Z_0}{2}
\ \rho_\mu^0 \rho^{0 \mu}\nonumber
\\&&-\frac{\delta Z_{\pm}}{2} \ \tilde\rho^+_{\mu\nu}\tilde\rho^{-
\mu\nu} +\left( \delta M^2+M^2 \delta Z_{\pm}\right) \ \rho_\mu^+ \rho^{- \mu}
\nonumber
\\ &&+ i\bar\psi
\gamma^\mu \ \delta Z_\Psi \partial_\mu \psi -\delta\lambda e \bar\psi \gamma^\mu
\ \frac{1+{\tau}^0}{2} \
 \psi \rho^0_\mu\nonumber\\
&&+ \left(\delta g + \frac{\delta Z_\alpha}{2} g_s\right) \bar\psi \gamma^\mu \
\frac{{\tau}^\alpha}{2} \
 \psi \ \rho^\alpha_\mu \nonumber \\
&&+ \frac{g_s}{2} \ \bar\psi\gamma^\mu  \left\{\delta
Z_\Psi,\frac{{\tau}^\alpha}{2}\right\}
 \psi \ \rho^\alpha_\mu  +\cdots,
 \label{lagrctl}
\end{eqnarray}
where we only display those counterterms explicitly which are relevant for the
subsequent discussion.
   All remaining terms are included in $\tilde{\cal L}_1$.

\begin{figure}
\begin{center}
\epsfig{file=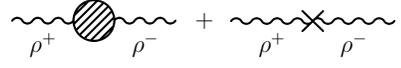,width=.6\linewidth}
\caption[]{\label{fse:fig} Self-energy diagrams.}
\end{center}
\end{figure}

   In order to establish relations among the renormalized coupling
constants pertaining to the Lagrangian of Eq.~(\ref{lagrmipctl}),
we analyze the renormalization of the self-energy of the charged
vector mesons and of the coupling constant $g_0$ using dimensional
regularization (with $n$ as a space-time dimension parameter) in
combination with the minimal subtraction (MS) scheme (for a
definition see, e.g., Ref.~\cite{Collins:xc}).
   However, our findings below do not depend on the choice of a specific
renormalization scheme.

    First we analyze the {\em divergent} parts of the self-energy of
the charged vector mesons to one-loop order (see Fig.~1). In particular, we are
interested in terms quadratic in the momenta. These terms consist of
counter\-term contributions,
\begin{equation}
\Pi_{\rm ct}^{\mu\nu} = \delta Z_{\pm} \left( p^2 g^{\mu\nu}-p^\mu p^\nu\right),
\label{rhosectc}
\end{equation}
and of contributions of one-loop diagrams,
\begin{equation}
\Pi_{\rm 1\,loop}^{\mu\nu}= A \ p^2 g^{\mu\nu}-B \ p^\mu p^\nu,
\label{rhosecloop}
\end{equation}
where $A$ and $B$ depend on the renormalized  parameters of the Lagrangian
\cite{thanks}. Perturbative renormalizability requires that the expressions in
Eqs.~(\ref{rhosectc}) and (\ref{rhosecloop}) cancel each other. This condition
implies that
\begin{equation}
A=B. \label{1stcond}
\end{equation}
Equation (\ref{1stcond}) is {\em not} automatically satisfied and imposes
constraints on the renormalized parameters of the Lagrangian.

    Next we analyze one-loop contributions to the $\rho^\pm \bar\psi\psi$ and $\rho^0
\bar\psi\psi$ vertices. These vertex functions (amputated Green's functions)
receive contributions from counter\-terms as well as one-loop diagrams. We again
require that the divergent parts have to cancel each other. From this condition
we determine the counterterm $\delta g$ as a function of the parameters of the
Lagrangian. As this counterterm contributes in both vertex functions we obtain
two expressions,
\begin{equation}
\delta g=\frac{1}{n-4} \ \phi_1(e,\kappa,m,M,c,g_s) \label{dg1}
\end{equation} and
\begin{equation}
\delta g=\frac{1}{n-4} \ \phi_2(e,\kappa,m,M,c,g_s). \label{dg2}
\end{equation}
The particular forms of the functions $\phi_1$ and $\phi_2$ have been determined
by calculating the loop diagrams of Fig.~2.
   Equations (\ref{dg1}) and (\ref{dg2}) result in
a condition for the renormalized parameters of the Lagrangian
\begin{equation}
\phi_1(e,\kappa,m,M,c,g_s)-\phi_2(e,\kappa,m,M,c,g_s)=0,
\label{constr2}
\end{equation}
which is not automatically satisfied. Solving Eqs.~(\ref{1stcond}) and
(\ref{constr2}) simultaneously, we obtain
\begin{equation}
c=e/g_s, \ \ \ \ \kappa=e. \label{relations}
\end{equation}
After taking the different normalization of Ref.~\cite{kappa} into
account, the second relation of Eq.~(\ref{relations}) agrees with
Ref.~\cite{Feuillat:2000ch}. Modeling the electromagnetic coupling
of the $\rho$ using vector meson dominance the same result has
been obtained in Ref.~\cite{Feuillat:2000ch} by arguing that the
electromagnetic self-mass should be finite.

   Since the magnetic moment of, say, the positively charged $\rho$ meson
reads
\begin{equation}
\vec{\mu}=\frac{e}{2M_{\rho^\pm}}\left(1+\frac{\kappa}{e}\right)\vec{S},
\label{mm}
\end{equation}
the second equality in Eq.~(\ref{relations}) leads to the gyromagnetic ratio
$g=2$ \cite{analogy,neutral}.
   Note that in the present case minimal coupling, as considered in
   Ref.~\cite{Hagen:1970wn},
would correspond to $\kappa =0$ yielding $g=1/s=1$ rather than
$g=2$ in agreement with Ref.~\cite{weinbergg2}. In
Ref.~\cite{weinbergg2}, $g=2$ was obtained by considering a
dispersion relation for forward Compton scattering, assuming that
the spin-dependent amplitude $f_{-}(\omega^2)$ vanishes at
infinity. Renormalizability is a matter of asymptotic behavior at
infinite momentum, so it may not be surprising that the
renormalizability (in the sense of EFT) is achieved for $g=2$.
However, we are not able to establish a connection between
perturbative renormalizability of the {\it low-energy} EFT and the
(high-energy) asymptotic behavior of the full amplitude.

Our derivation of Eq.~(\ref{relations}) and therefore the value
$g=2$ hold independently whether or not pion degrees of freedom
are included explicitly. As the pionless effective theory can be
obtained by integrating out the pion fields, the matching
condition together with Eq.~(\ref{relations}) requires that the
pion loop corrections to the gyromagnetic ratio vanish. We
emphasize that this argument only holds if there exist
self-consistent effective field theories both with and without
explicit pion degrees of freedom.

Calculating the propagator poles at tree level and using Eq.~(\ref{relations}),
we find
\begin{equation}
M_{\rho^\pm}^2=M_{\rho^0}^2 \ \left( 1-\frac{e^2}{g_s^2}\right).
\label{massrelation}
\end{equation}
Substituting numerical values for $g_s$ [estimated from the KSRF relation
\cite{Kawarabayashi:1966kd,Riazuddin:sw}, $g_s^2=M_\rho^2/(2F_\pi^2)$, with
$F_\pi=92.4$ MeV and $M_\rho=769$ MeV] and $e$ we obtain as a prediction for the
mass difference
\begin{equation}
M_{\rho^0}-M_{\rho^\pm}\sim 1 \ {\rm MeV} \label{massdifference}
\end{equation}
which has to be compared with the present PDG average of $(0.7\pm 0.7)$ MeV
\cite{Eidelman:2004wy}.

\begin{figure}
\begin{center}
\epsfig{file=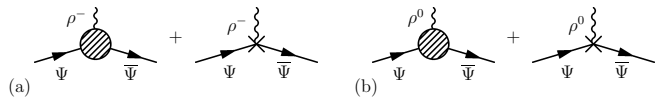,width=\linewidth}
\caption[]{\label{vertex:fig} Vertex diagrams.}
\end{center}
\end{figure}

   To summarize, we have considered quantum electrodynamics of
$\rho$ mesons as an effective field theory described by the most
general Lagrangian consistent with all symmetries of the theory.
    The self-consistency condition of this theory (imposed by the renormalization
procedure) constrains the parameters of the Lagrangian. In
particular, instead of depending on a free parameter, the
gyromagnetic ratio of the charged vector meson is $g=2$ (up to
loop corrections). The mixing parameter of the photon and the
neutral vector meson is equal to the ratio of electromagnetic and
strong coupling constants, leading to $M_{\rho^0}-M_{\rho^\pm}\sim
1 \ {\rm MeV}$ at tree order.

\begin{acknowledgments}
The work of J.~Gegelia and M.~R.~Schindler was supported by the
Deutsche Forschungsgemeinschaft (SFB 443).

\end{acknowledgments}


\begin{thebibliography}{100}
\bibitem{ren}
Our analysis of the renormalizability of the considered EFT is by no means
complete. Our results are obtained by considering some of the necessary
conditions of the existence of a self-consistent effective field theory. Although
we believe that, in the sense of renormalizability, there exists a consistent EFT
of nucleons, photons, pions, and vector-mesons, to the best of our knowledge this
has never been proven.
\bibitem{f1}
We only analyze interaction terms with dimensionless coupling
constants.
   The explicit inclusion of non-renormalizable interaction terms, included in ${\cal L}_1$,
is not necessary for the purpose of this work. All terms
containing pion fields are also included in ${\cal L}_1$.
\bibitem{analogy}
It is worth noting a close analogy with the Salam-Weinberg model, where $g=2$ for
the W bosons and the relation between the mixing parameter ($\sin\theta_W$) and
the coupling constants are obtained as consequences of the
$\mbox{SU(2)}\times\mbox{ U(1)}$ gauge symmetry.
\bibitem{neutral}
Note that invariance under charge conjugation excludes a $\gamma\rho^0\rho^0$
vertex.
\end{thebibliography}
\end{document}